\documentclass[twocolumn,prb]{revtex4}
\usepackage{amsfonts}
\usepackage[T1]{fontenc}
\usepackage{amsmath,amsbsy,amssymb,graphicx}
\usepackage{times}

\begin{document}

\title{ Topological Micro-Electro-Mechanical Systems}
\author{Motohiko Ezawa}
\affiliation{Department of Applied Physics, University of Tokyo, Hongo 7-3-1, 113-8656,
Japan}

\begin{abstract}
We explore the topological aspect of dynamics in a micro-electro-mechanical
system (MEMS), which is a combination of an electric-circuit system and a
mass-spring system. A simplest example is a sequential chain of capacitors
and springs. It is shown that such a chain exhibits novel topological
dynamics with respect to its oscillation modes. On one hand, when it
undergoes free oscillation, the system is governed by the
Su-Schrieffer-Heeger model, and the topological charge is given by a winding
number. Topological and trivial phases are differentiated by measuring the
dynamics of the outermost plate. On the other hand, when it undergoes
periodical motion in time, the system is governed by an
inversion-symmetric-trimer model, and the topological phases are
characterized by an inversion-symmetry indicator. There emerge topological
edge states, which are well signaled by measuring
electromechanical-impedance resonance. Our results will open an attractive
field of topological MEMS.
\end{abstract}

\maketitle

\textbf{Introduction}: Topological physics is one of the most important
developments made over the past few decades\cite{Qi}. It has been
investigated mainly in the context of electron systems in solid. However, it
is now expanded to artificial topological systems such as photonic\cite{Lu,Ozawa,KhaniPhoto}, 
acoustic\cite{Xiao,TopoAco,Mous,GarciaAco,Xue,Ni},
mechanical\cite{Prodan,Lubensky,Nash,Sus,Kariyado,Takahashi,Mee,Abba,Mat}
and electric-circuit\cite%
{TECNature,ComPhys,Hel,Research,Zhao,EzawaTEC,Garcia,Hofmann,EzawaNH,EzawaSkin,EzawaMajo,EzawaTQC}
systems. For example, in electric circuits, the circuit Laplacian is
identified with the Hamiltonian of the system, and the topological edge
states are well signaled by impedance resonance\cite{TECNature,ComPhys}.

Micro-electro-mechanical systems (MEMS) are combined systems of electric
circuits and mechanical systems, which is a key technology of current
industry. Actuators convert a mechanical motion to an electric signal. Among
them, the parallel-plate electrostatic actuator is a simplest example, where
the plates of a capacitor are connected with a spring. Since the energy
stored in a capacitor depends on the separation between two plates
constituting the capacitor, the mechanical motion of the parallel-plate is
transformed into an electric signal. It is a basis of microphone.

In this work, we reveal the presence of a novel topological structure in the
dynamics of MEMS. We propose a simplest model consisting of a sequential
connection of capacitors and springs. It is intriguing that such a system
exhibits entirely different dynamical behaviors as a function of a system
parameter. We explain it in terms of topological phase transitions described
by the Su-Schrieffer-Heeger model or the trimer model depending on free or
forced oscillation.

First, we study a free oscillation, which occurs without applying external
force and voltage. It is shown that the oscillation dynamics is described by
the Su-Schrieffer-Heeger model (SSH) model, where the topological charge is
the winding number. Topological and trivial phases are differentiated
experimentally by measuring the dynamics of the outermost plate. Second, we
study a forced oscillation, driving the system by external force or voltage.
It is shown that the oscillation dynamics is described by a trimer model
with inversion symmetry. The topological charge is given with the use of an
inversion-symmetry indicator, which counts the emergent topological edge
states. The topological edge states are well observed experimentally by
measuring electromechanical-impedance resonance. We show numerically that
these dynamical properties are robust against randomness and damping. It is
interesting that two different topological systems emerge by changing the
experimental setup in a single MEMS.

\begin{figure}[t]
\centerline{\includegraphics[width=0.48\textwidth]{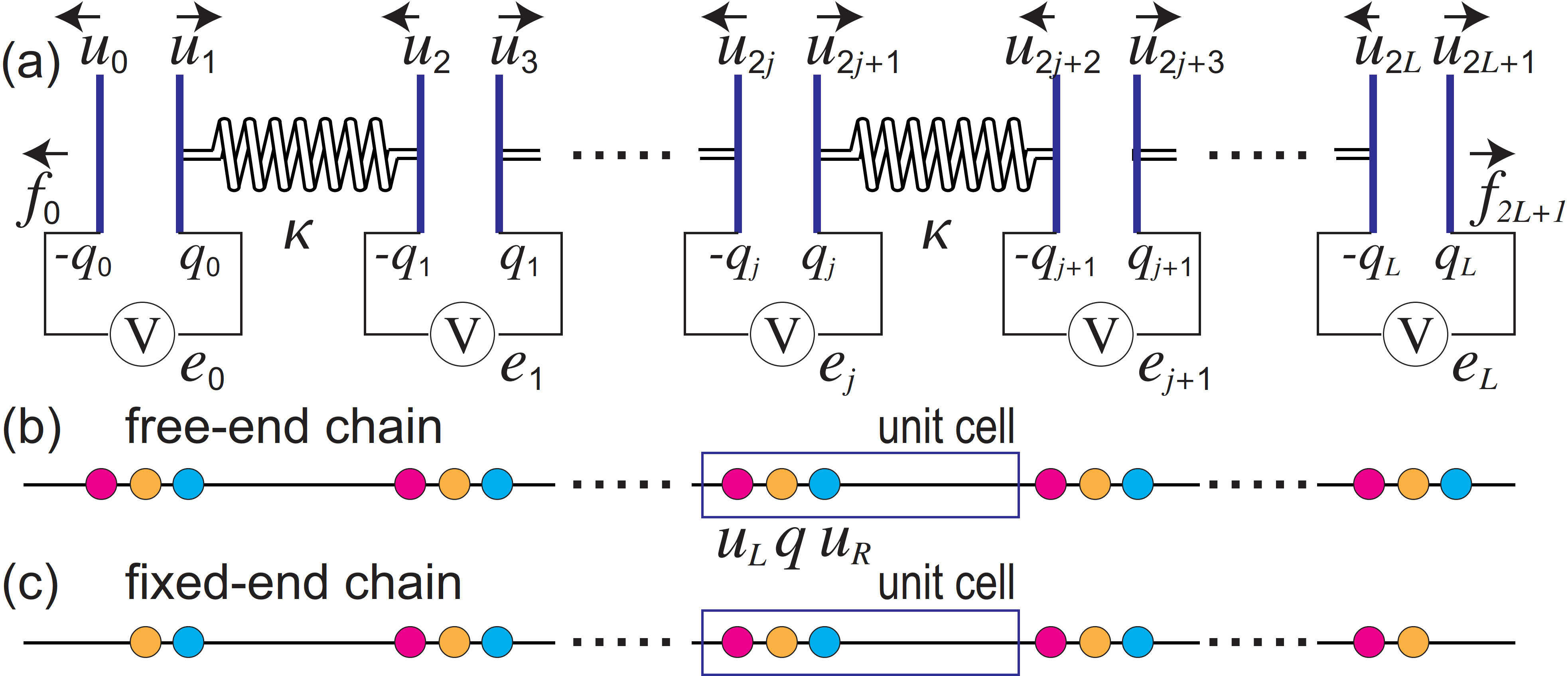}}
\caption{(a) Illustration of MEMS. Parallel metallic plates of capacitors
can move, where adjacent plates are connected by insulating springs. The
displacement $u_{j}$ of the $j$th plate is measured from its equilibrium
position, while the charge $q_{j}$ is measured from the equilibrium charge 
$Q_{0}$ of the capacitor. Each capacitor is connected to a battery in order
to charge up. One-dimensional tight-binding model (b) for a free-end chain
and (c) for a fix-end chain, where the unit cell contains three sites marked
in magenta, orange and cyan. }
\label{FigMEMS}
\end{figure}

\textbf{Micro-electro-mechanical systems (MEMS)}: We consider a system where
capacitors and springs are connected sequentially as shown in Fig.\ref{FigMEMS}, 
where the spring has elastic constant $\kappa $. A capacitor is
made of two plates, where each plate has mass $m$. Each plate is bound to
its equilibrium position by a spring with elastic constant $\kappa _{0}$. We
focus on the $j$th capacitor. Let the displacement of the left (right)
parallel plate measured from its equilibrium position be $u_{2j}$ and 
$u_{2j+1}$, where we use a convention such that $u_{2j}>0$ ($u_{2j+1}>0$) for
the leftward (rightward) displacement. The actual distance between the two
plates is $\ell _{j}^{\text{cap}}=X_{\text{cap}}+u_{2j+1}+u_{2j}$ 
with $X_{\text{cap}}$ being the equilibrium distance. On the other hand, the length
of the spring is $\ell _{j}^{\text{spr}}=X_{\text{spr}}-u_{2j}-u_{2j-1}$
with $X_{\text{spr}}$ being the equilibrium length. The springs are
insulating, where there is no charge transfer between adjacent capacitors.
Finally, $q_{j}$ is the charge deviation from the equilibrium charge $Q_{0}$
in the $j$th capacitor. We attach a battery for each capacitor in order to
charge up $Q_{0}$. The system shows a coupled oscillation. We analyze the
dynamics of the system as a function of $Q_{0}$.

The Lagrangian of the system is given by 
\begin{align}
L& =\sum_{j}\left[ \frac{m}{2}(v_{2j-1}^{2}+v_{2j}^{2})-\frac{\kappa _{0}}{2}\left( u_{2j-1}^{2}+u_{2j}^{2}\right) \right.  \notag \\
& \left. -\frac{\kappa }{2}\left( X_{\text{spr}}-u_{2j-1}-u_{2j}\right) ^{2}-\frac{\left( Q_{0}+q_{j}\right) ^{2}}{2C(u)}\right] ,
\end{align}%
where $v_{j}=du_{j}/dt$ is the velocity of the plate, while 
$C(u)=\varepsilon _{0}S/(X_{\text{cap}}+u_{2j}+u_{2j+1})$ is the capacitance
with area $S$ and permittivity $\varepsilon _{0}$. The Euler-Lagrange
equations are 
\begin{align}
F_{0}+f_{j}& =\frac{d}{dt}\left( \frac{\partial L}{\partial v_{j}}\right) -\frac{\partial L}{\partial u_{j}}+\frac{\partial R}{\partial v_{j}},
\label{EuLaA} \\
E_{0}+e_{j}& =\frac{d}{dt}\left( \frac{\partial L}{\partial I_{j}}\right) -\frac{\partial L}{\partial q_{j}},  \label{EuLaB}
\end{align}%
where $f_{j}$ is the force acting on the $j$th plate of the capacitor
measured from the equilibrium force $F_{0}$, and $e_{j}$ is the voltage
between the two plates of the $j$th capacitor measured from the equilibrium
voltage $E_{0}$. $I_{j}=dq_{j}/dt$ is the current flowing the $j$th
capacitor, which is absent in the Lagrangian and we have $\partial
L/\partial I_{j}=0$. We have introduced the Rayleigh dissipation function 
$R=\sum_{j}\frac{\gamma }{2}v_{j}^{2}$, with $\gamma $ being the damping
factor. The equilibrium conditions read $\kappa X_{\text{spr}}=Q_{0}^{2}/(2\varepsilon _{0}S)$ 
and $E_{0}=X_{\text{cap}}Q_{0}/(\varepsilon _{0}S)$, where we have assumed $F_{0}=0$.

We explicitly write down the Euler-Lagrange equations (\ref{EuLaA}) and (\ref{EuLaB}), 
and expand them to the linear order in $u_{2j-1}$, $u_{2j}$ and 
$q_{j}$. Making the Fourier transformation from time $t$ to frequency $\omega$, 
we obtain for the $j$th unit cell (see Fig.\ref{FigMEMS}) that 
\begin{align}
f_{2j}& =-M_{0}u_{2j}+\kappa u_{2j-1}+\frac{Q_{0}}{\varepsilon _{0}S}q_{j},
\label{EqA} \\
e_{j}& =\frac{Q_{0}}{\varepsilon _{0}S}\left( u_{2j+1}+u_{2j}\right) +\frac{X_{\text{cap}}}{\varepsilon _{0}S}q_{j},  \label{EqB} \\
f_{2j+1}& =-M_{0}u_{2j+1}+\kappa u_{2j+2}+\frac{Q_{0}}{\varepsilon _{0}S}q_{j},  \label{EqC}
\end{align}%
where $M_{0}=m\omega ^{2}-\kappa -\kappa _{0}-i\omega \gamma $.

We redefine the charge as $\hat{q}_{j}\equiv X_{\text{cap}}q_{j}/Q_{0}$ so
that it has the dimension of length, and the voltage as 
$\hat{e}_{j}=Q_{0}e_{j}/X_{\text{cap}}$ so that it has the dimension of force. For a
spatially periodic system, by making the Fourier transformation from the
site index $j$ to momentum $k$ ($-\pi <k\leq \pi $), a set of equations 
(\ref{EqA}), (\ref{EqB}) and (\ref{EqC}) is transformed into 
\begin{equation}
\left( 
\begin{array}{c}
f_{\text{L}} \\ 
\hat{e} \\ 
f_{\text{R}}%
\end{array}%
\right) =\left( 
\begin{array}{ccc}
-M_{0} & Q & \kappa e^{-ik} \\ 
Q & Q & Q \\ 
\kappa e^{ik} & Q & -M_{0}%
\end{array}%
\right) \left( 
\begin{array}{c}
u_{\text{L}} \\ 
\hat{q} \\ 
u_{\text{R}}%
\end{array}%
\right) ,  \label{EqMEMS}
\end{equation}%
where $Q=Q_{0}^{2}/\left( X_{\text{cap}}\varepsilon _{0}S\right) $; 
$u_{\text{L}}(k)$, $\hat{q}\left( k\right) $ and $u_{\text{R}}(k)$ are the
Fourier components of $u_{2j}$, $q_{j}$ and $u_{2j+1}$, while $f_{\text{L}}(k)$, 
$\hat{e}\left( k\right) $ and $f_{\text{R}}(k)$ are the Fourier
components of $f_{2j}$ $\hat{e}_{j}$, and $f_{2j+1}$.

We represent Eq.(\ref{EqMEMS}) in the form of 
\begin{equation}
\Phi _{fe}=J\Phi _{uq},  \label{Admi}
\end{equation}%
where $J$ is a $3\times 3$ matrix. We call $J$ the MEMS Laplacian as in the
case of the circuit Laplacian in electric circuits, where $J$ has the
dimension of spring constant.

A finite length chain may have either free ends or fixed ends. In the
present problem, the free-end (fixed-end) oscillation occurs when a chain is
terminated at the positions of the capacitor planes $u_{L}$ and $u_{R}$
(charge $q$) as in Fig.\ref{FigMEMS}(b) [(c)]. Note that the free-end chain
respects the unit cell, but the fixed-end chain does not.

\begin{figure}[t]
\centerline{\includegraphics[width=0.48\textwidth]{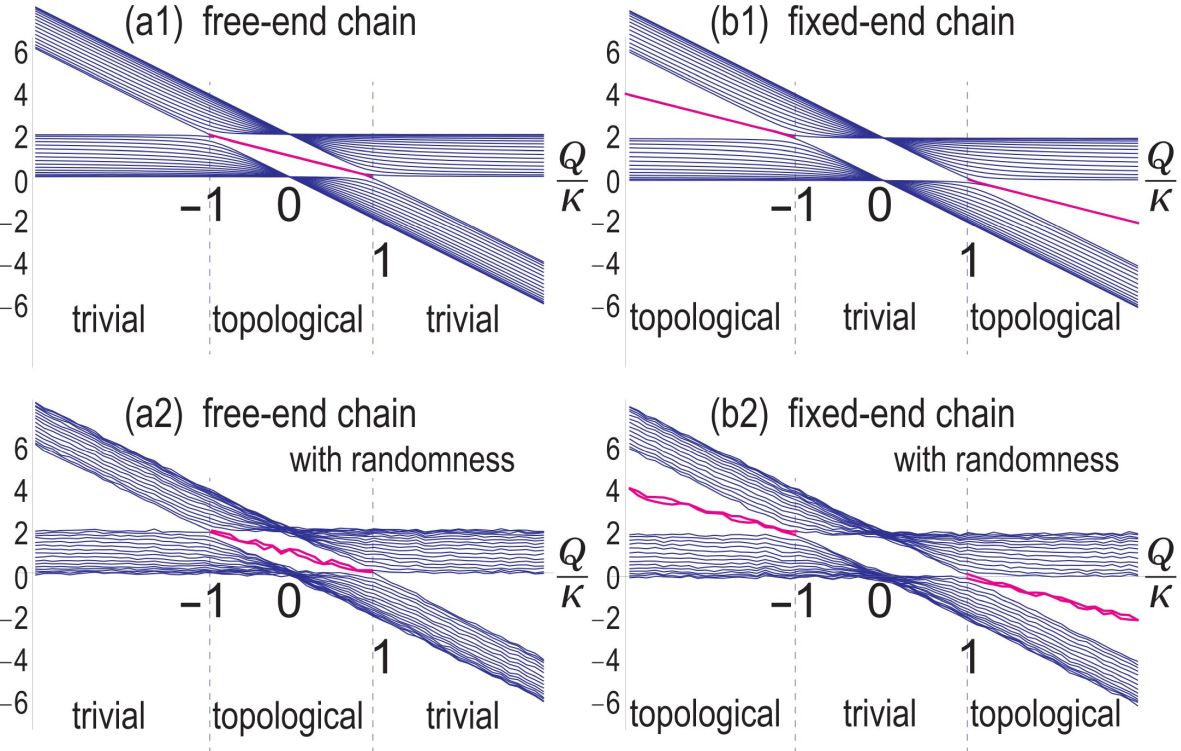}}
\caption{(a1) Eigen spectra of the part ($K-\protect\kappa _{0}I_{2}$) of
the dynamical matrix for a free-end chain and (b1) that for a fix-end chain
in the case of free oscillation. The emergence of topological edge states
marked in magenta differentiates the topological and the trivial phases. The
horizontal axis is $Q/\protect\kappa $. Only the region $Q/\protect\kappa >0$
is physical. We have set $\protect\gamma =0$. (a2) and (b2) are those with
inclusion of randomness distributing uniformly from $-0.1\protect\kappa$ to 
$0.1\protect\kappa$ in the spring constants $\protect\kappa$, $\protect\kappa_0$ and the charge $Q_0$. }
\label{FigSSHRibbon}
\end{figure}

\textbf{Free oscillation:} First, we study the case of free oscillation in
the absence of external force $f$ and voltage $e$, $\Phi _{fe}=0$, 
where Eq.(\ref{Admi}) becomes $J\Phi _{xq}=0$. The charge is solved as 
$\hat{q}=-\left( u_{\text{L}}+u_{\text{R}}\right) $. We use it to eliminate $\hat{q}$
from Eq.(\ref{EqMEMS}), and obtain the kinetic equation,%
\begin{equation}
m\frac{d^{2}}{dt^{2}}+\gamma \frac{d}{dt}\left( 
\begin{array}{c}
u_{\text{L}} \\ 
u_{\text{R}}%
\end{array}%
\right) =-K\left( 
\begin{array}{c}
u_{\text{L}} \\ 
u_{\text{R}}%
\end{array}%
\right) ,
\end{equation}%
with the dynamical matrix $K=(\kappa +\kappa _{0}-Q)I_{2}-H_{\text{SSH}}$,
where $H_{\text{SSH}}$ is the SSH model,%
\begin{equation}
H_{\text{SSH}}=\left( 
\begin{array}{cc}
0 & Q-\kappa e^{-ik} \\ 
Q\mathcal{-}\kappa e^{ik} & 0%
\end{array}%
\right) .
\end{equation}%
The spring constant $\kappa $ becomes $\kappa +\kappa _{0}-Q$ for $H_{\text{SSH}}=0$. 
This is known as the soft spring effect in the context of MEMS. We
can tune $\kappa _{0}$ to make all the eigenvalues of $K$ positive.

The phase diagram is constructed as a function of the dimensionless
parameter $Q/\kappa $. The SSH model is topological (trivial) for $|Q/\kappa
|<1$ and trivial (topological) $|Q/\kappa |>1$, as is clear for a free-end
(fixed-end) chain in Fig.\ref{FigSSHRibbon}(a1) and (b1). The topological
number is the winding number given by $W=\frac{1}{2\pi i}\int \left\langle
\psi \right\vert \partial _{k}\left\vert \psi \right\rangle dk$, where $\psi 
$ is the eigenfunction of $H_{\text{SSH}}$. It is $1$ for the topological
phase and $0$ for the trivial phase. It defines chiral-symmetry-protected
topological (chiral-SPT) phases.

\begin{figure}[t]
\centerline{\includegraphics[width=0.48\textwidth]{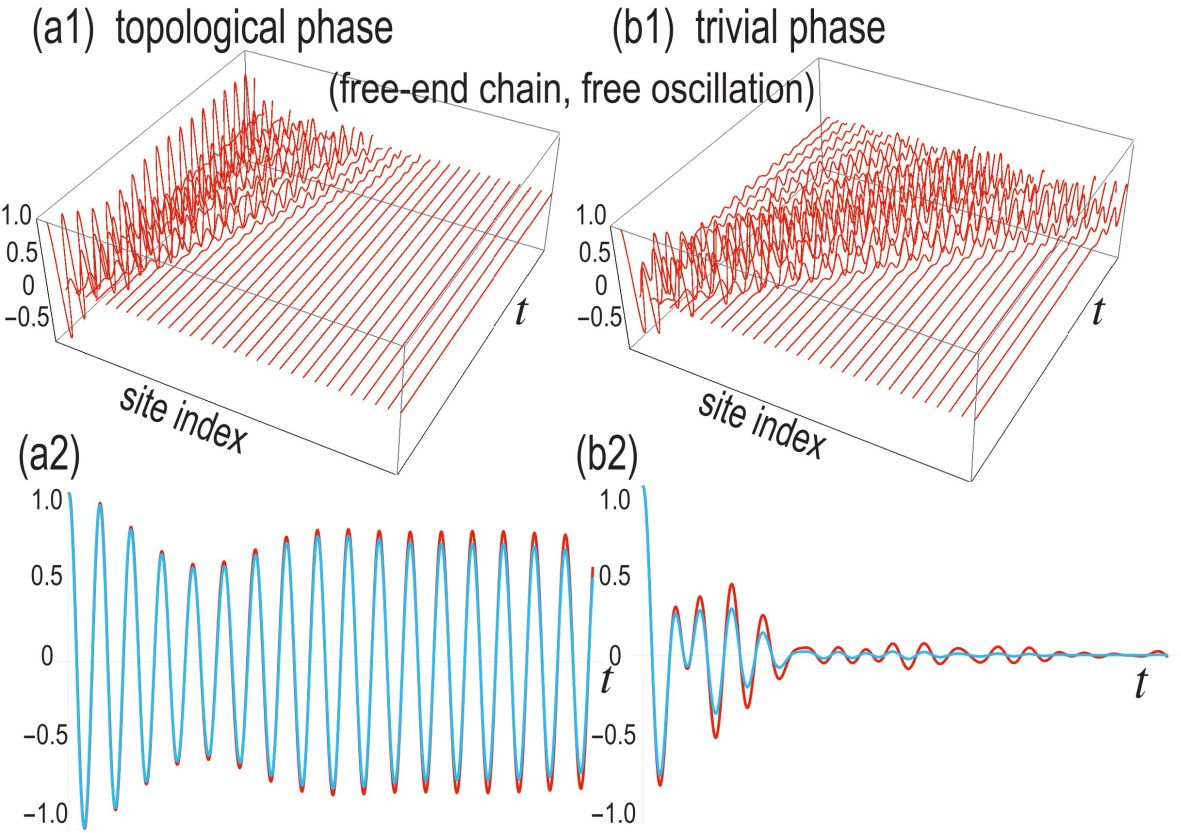}}
\caption{(a1) Time evolution of the plate displacement $u_{j}(t)$ in the
topological phase with $Q/\protect\kappa =0.5$ and (b1) that in the trivial
phase with $Q/\protect\kappa =1.5$ for a free-end chain in the case of free
oscillation. (a2) The time evolution of the outer-most plate displacement 
$u_{0}(t)$ continues to oscillate with a finite amplitude in the topological
phase, while (b2) it damps rapidly in the trivial phase. We have set 
$\protect\kappa _{0}=4$. In (a2) and (b2), red curves are for no damping 
($\protect\gamma =0$), which are overwritten by cyan curves with damping 
($\protect\gamma =0.1\protect\kappa /\protect\omega _{0}$). 
Note that $\protect\gamma \sim 0.01\protect\kappa /\protect\omega _{0}$ in actual samples,
whose effects are negligible.}
\label{FigSSH}
\end{figure}

We show the time evolution of the plate displacement $u_{j}(t)$ in Fig.\ref{FigSSH}(a1) and (b1) for all $j$. 
We impose the initial condition such that
only the outermost plate at $j=0$ is displaced with the initial zero
velocity. An oscillation propagates into the bulk in both the topological
and the trivial phases. However, there is a significant difference. In the
trivial phase the oscillation propagates rapidly into the whole bulk, but in
the topological phase the oscillation is almost localized at the outermost
plate. In particular, the outermost plate exhibits a salient oscillation
dynamics. The outermost plate continues to oscillate with a finite amplitude
in the topological phase, while it rapidly damps in the trivial phase. This
is because its oscillation mode is almost isolated from (embedded in) the
bulk spectrum in the topological (trivial) phase. We also show the time
evolution of the plate in the presence of the damping term $\gamma $ in Fig.\ref{FigSSH}(a2) and (b2). 
The oscillation damps but its effect is very
small. Indeed, the damping is negligible since $\gamma \sim 10^{-2}\kappa
/\omega _{0}$ in actual samples.

\begin{figure}[t]
\centerline{\includegraphics[width=0.48\textwidth]{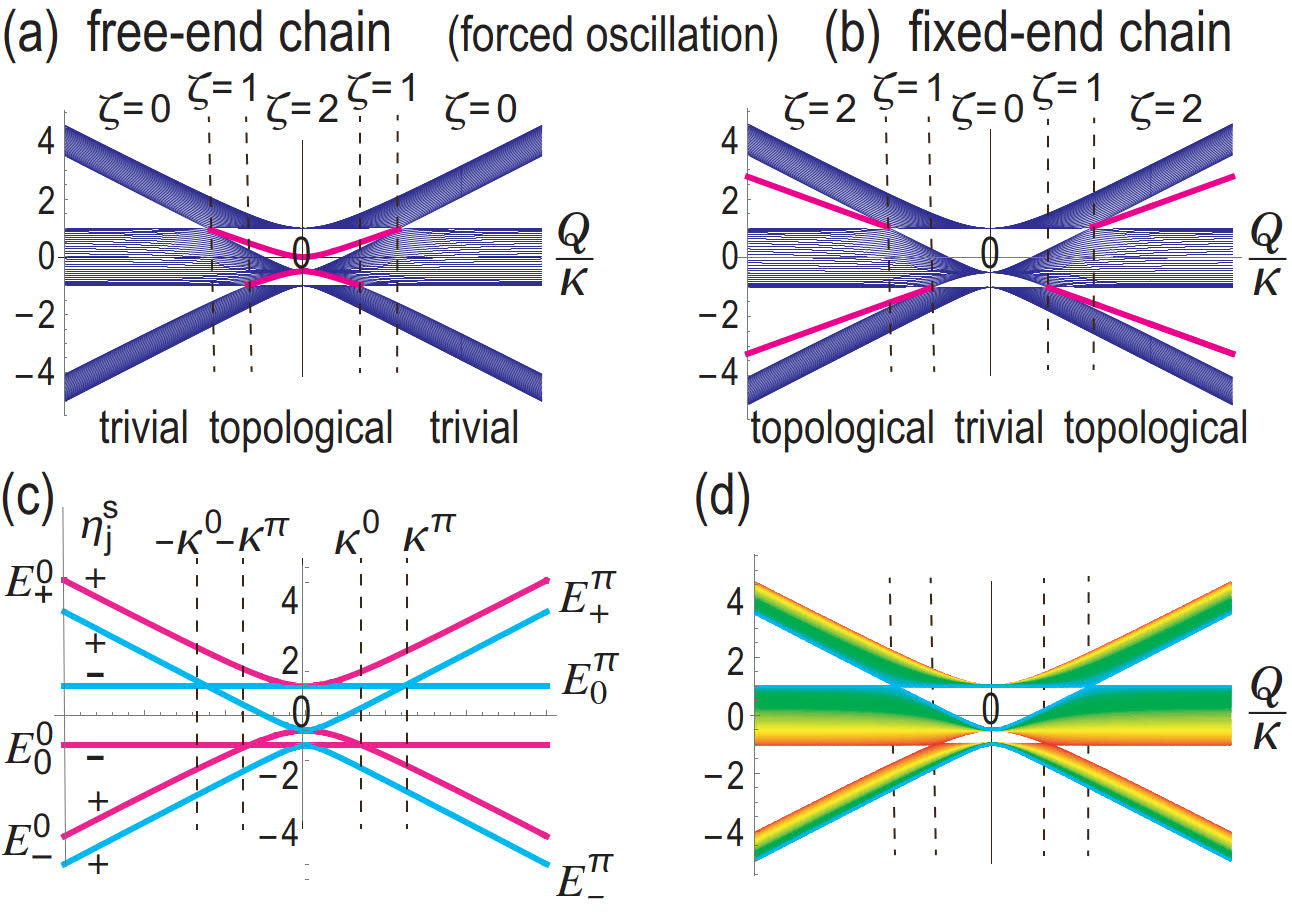}}
\caption{(a) Admittance spectrum for a finite free-end chain and (b) that
for a finite fixed-end chain in the case of forced oscillation. We have set $M=1.5\protect\kappa $. 
The horizontal axis is $Q/\protect\kappa $, while the
vertical axis is the admittance of the MEMS Laplacian. Topological edge
states are marked in magenta. (c) Analytically obtained bulk admittance 
$E_{j}^{k}$ for $k=0$ and $\protect\pi $. The signs $\protect\eta _{j}^{k}=+$
and $-$ indicate the sign of the inversion-symmetry indicator. (d)
Admittance spectrum for various $k$, where the color is ranging from $k=0$
marked in magenta to $k=\protect\pi $ marked in cyan. Only the region $Q/\protect\kappa >0$ is physical. 
We have set $\protect\gamma =0$.}
\label{FigRibbon}
\end{figure}

A comment is in order with respect to chiral symmetry. In actual samples,
chiral symmetry is broken by randomness in the spring constants $\kappa $, 
$\kappa _{0}$ and the charge $Q_0$. Hence, strictly speaking, the system has
no SPT phases. Nevertheless, even if we introduce fluctuations as much as
10\% in these material constants, the topological edge states are found to
be robust as shown in Fig.\ref{FigSSHRibbon}(a2) and (b2), although the
admittance of the edge states slightly shifts. Consequently, the topological
dynamics of MEMS is experimentally observable in actual samples with
randomness.

\textbf{Forced oscillation:} We next investigate a forced oscillation driven
by external force or voltage periodic in time. We introduce a new quantity 
$H_{\text{tri}}$ by $H_{\text{tri}}=J+M_{0}I_{3}$, 
\begin{equation}
H_{\text{tri}}=\left( 
\begin{array}{ccc}
0 & Q & \kappa e^{-ik} \\ 
Q & M-\kappa & Q \\ 
\kappa e^{ik} & Q & 0%
\end{array}%
\right) .  \label{EqM}
\end{equation}%
with $M=m\omega ^{2}-\kappa _{0}+Q-i\omega \gamma $. Here, we have made
explicit how $H_{\text{tri}}$ depends on $Q$ and $\kappa $. It is a key
observation that $H_{\text{tri}}$ is identical to the Hamiltonian of a
trimer model\cite{Agar,Jin,Alva} on one-dimensional lattice, where the unit
cell contains three sites, as illustrated in Fig.\ref{FigMEMS}(b) and (c).

The trimer-model Hamiltonian has inversion symmetry $PH_{\text{tri}}\left(
k\right) P^{-1}=H_{\text{tri}}\left( -k\right) $, where 
\begin{equation}
P=\left( 
\begin{array}{ccc}
0 & 0 & 1 \\ 
0 & 1 & 0 \\ 
1 & 0 & 0%
\end{array}%
\right) .
\end{equation}%
It plays a key role to define inversion-SPT phases.

There are two inversion-symmetric momenta $k=0,\pi $. At these points, the
eigenenergies $E_{j}^{k}$ and eigenfunctions $\psi _{j}^{k}$ are
analytically obtained, 
\begin{align}
E_{0}^{0}& =-\kappa ,\quad E_{\pm }^{0}=\frac{M}{2}\pm \frac{\sqrt{\left(
M-2\kappa \right) ^{2}+8Q^{2}}}{2}, \\
E_{0}^{\pi }& =\kappa ,\quad E_{\pm }^{\pi }=\frac{M}{2}-\kappa \pm \frac{\sqrt{M^{2}+8Q^{2}}}{2}, \\
\psi _{0}^{k}& =(-1,0,1)/\sqrt{2},\quad \psi _{\pm }^{k}=(1,F_{\pm }^{k},1)/\sqrt{2+F_{\pm }^{2}},
\end{align}%
with%
\begin{align}
F_{\pm }^{0}& =\{M-2\kappa \pm \sqrt{\left( M-2\kappa \right) ^{2}+8Q^{2}}\}/\left( 2Q\right) , \\
F_{\pm }^{\pi }& =\{M\pm \sqrt{M^{2}+8Q^{2}}\}/\left( 2Q\right) .
\end{align}%
We show the spectrum $E_{j}^{k}(Q)$ in Fig.\ref{FigRibbon}(c), where the
horizontal axis is a dimensionless quantity $Q/\kappa $. By solving $E_{\pm
}^{0}(Q)=E_{0}^{0}(Q)$, we obtain a gap closing point $\pm Q^{0}$ with 
$Q^{0}=\sqrt{M\kappa }$. By solving $E_{\pm }^{\pi }(Q)=E_{0}^{\pi }(Q)$, we
obtain two gap closing points $\pm Q^{\pi }$ with $Q^{\pi }=\sqrt{\kappa
\left( 2\kappa -M\right) }$. They are the endpoints of the magenta curves in
Fig.\ref{FigRibbon}(a) and (b).

\textit{Topological edge states:} The admittance is the eigenvalue of the
MEMS Laplacian $J$. We show the admittance spectrum of a finite length chain
as a function of $Q/\kappa $ in Fig.\ref{FigRibbon}(a) and (b). The magenta
curves express edge states isolated from the bulk spectra at each $\kappa $.
As suggested by the bulk-edge correspondence, they would be topological edge
states at for $\left\vert Q\right\vert <Q^{0} $ for the lower band and 
$\left\vert Q\right\vert <Q^{\pi }$ for the upper band in Fig.\ref{FigRibbon}(a), 
and for $\left\vert Q\right\vert >Q^{0}$ for the lower band and 
$\left\vert Q\right\vert >Q^{\pi }$ for the upper band in Fig.\ref{FigRibbon}(b). 
To confirm this observation, we define the topological charge counting
the number of magenta curves in what follows. Then, the phase diagram is
given in term of a dimensionless parameter $Q/\kappa $ as in Fig.\ref{FigRibbon}(a) and (b).

\textit{Inversion-symmetry indicator:} One dimensional system with inversion
symmetry is classified\cite{InvClass,Shiozaki} into the class $\mathbb{Z}$. 
See Table VII class A with $\delta =1$ in Ref.\cite{Shiozaki}.
Symmetry indicators are good candidates to characterize SPT phases\cite{Wang,Po,Song,Khalaf,PoJ,Slager}. 
The inversion symmetry operator acts on
the eigenfunctions as $P\psi _{0}^{k}=\eta _{0}^{k}\psi _{0}^{k}$, 
$P\psi_{\pm }^{k}\left( k\right) =\eta _{\pm }^{k}\psi _{\pm }^{k}\left( k\right) $, 
with $\eta _{0}^{k}=-1$ and $\eta _{\pm }^{k}=1$, where $k=0,\pi $. We
define the inversion-symmetry indicator for the occupied band by $\zeta
^{k}=\sum_{j}\eta _{j}^{k}$, where the sum over $j=0,\pm $ is taken for all
occupied bands. See Fig.\ref{FigRibbon}(c), where $\eta _{j}^{k}=\pm $ is
assigned to each curve $E_{j}^{k}(\kappa )$.

\begin{figure}[t]
\centerline{\includegraphics[width=0.48\textwidth]{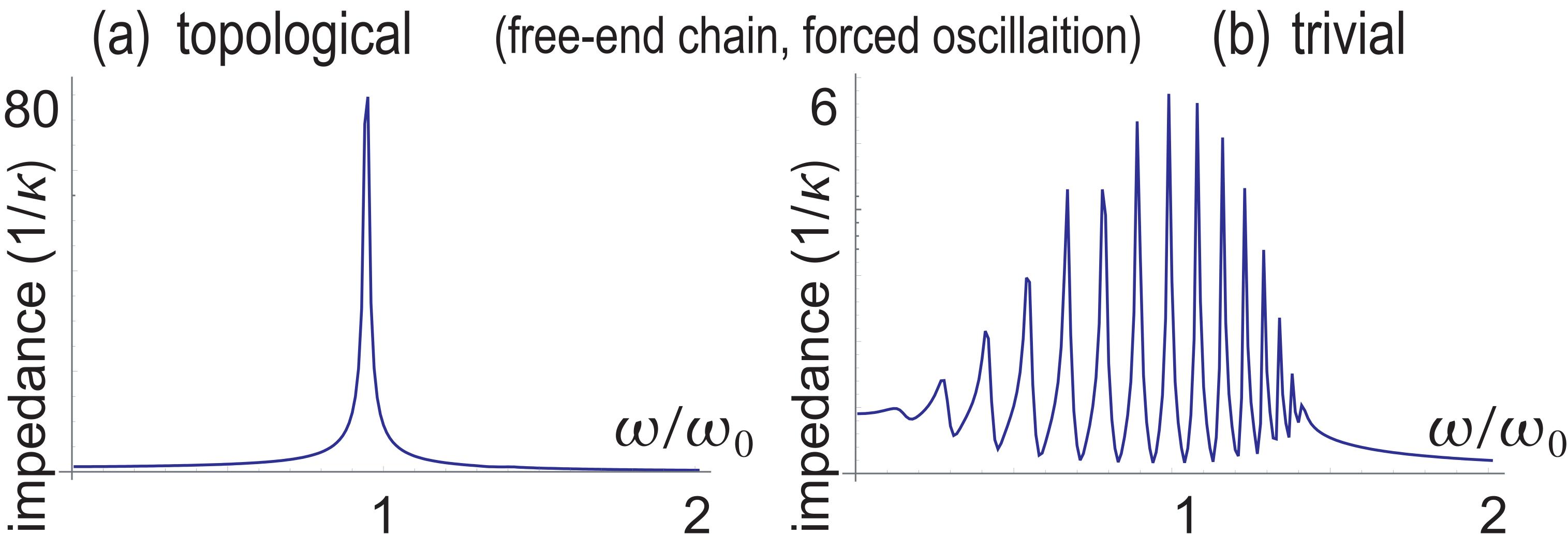}}
\caption{(a) Absolute value of electromechanical impedance in the
topological phase with $Q=0.25$ and (b) that in the trivial phase with $Q=2$
in a free-end chain in the case of forced oscillation. It depends
sensitively on the damping factor $\protect\gamma$. Here, we have used a
realistic value, $\protect\gamma =0.01\protect\kappa /\protect\omega_0 $. It
diverges for $\protect\gamma =0$. The horizontal axis is the driving
frequency $\protect\omega /\protect\omega _{0}$ with $\protect\omega _{0}=\protect\sqrt{\protect\kappa /m}$. 
We have set $Q-\protect\kappa _{0}=0.5\protect\kappa $. We have used a finite chain containing 48 sites.}
\label{FigImpe}
\end{figure}

We now define the topological number $\zeta $ with the use of the
inversion-symmetry indicators so that $\zeta $ is nonzero for the
topological phase and $\zeta =0$ for the trivial phase. (i) For a free-end
chain, we define the number $\zeta _{\text{low}}=|\zeta ^{\pi }-\zeta
^{0}|/2 $ for the lowest spectrum. We obtain $\zeta _{\text{low}}=1$ for 
$\left\vert Q\right\vert <Q^{0}$, while we obtain $\zeta _{\text{low}}=0$ for 
$\left\vert Q\right\vert >Q^{0}$, which counts the number of the lower edge
states. In the similar way, the number of the upper edge states is counted
by the number $\zeta _{\text{high}}=|\zeta ^{\pi }-\zeta ^{0}|/2$ for the
highest spectrum, that is, $\zeta _{\text{high}}=1$ for $\left\vert
Q\right\vert <Q^{\pi }$ and $\zeta _{\text{high}}=0$ for $\left\vert
Q\right\vert >Q^{\pi }$. The topological number is given by 
$\zeta =\zeta _{\text{low}}+\zeta _{\text{high}}$, which agrees with the number of
topological edge states shown in Fig.\ref{FigRibbon}. (ii) For a fixed-end
chain, it is enough to define the topological number $\zeta $ in terms of 
$\zeta _{\text{low}}$ and $\zeta _{\text{high}}$ given by $|\zeta ^{\pi}+\zeta ^{0}|/2$.

\textit{Electromechanical impedance:} Impedance resonance is a good signal
to detect topological edge states in electric circuits\cite{TECNature,ComPhys,EzawaTEC}. 
We now show that the electromechanical
impedance well signals the emergence of the topological edge states. We
define the electromechanical impedance by the inverse of the MEMS Laplacian,
i.e., by $Z=J^{-1}=G$, where $G$ is the Green function. It is obtained by
measuring the displacement $u$ and the charge $q$ induced under the
application of the force $f$ or voltage $e$ by way of the relation (\ref{Admi}) or $\Phi _{uq}=Z\Phi _{fe}$. 

For a semi-infinite free-end chain, the left-upper most component of the
Green function is $G_{00}$, which is expressed in terms of a continued
fraction $G_{00}=\mathcal{G}/\kappa ^{2}$ as%
\begin{equation}
\mathcal{G}=\frac{1}{-M_{0}-\frac{Q^{2}}{Q-\frac{Q^{2}}{-M_{0}-\frac{\kappa
^{2}}{-M_{0}-\cdots }}}}=\frac{1}{-M_{0}-\frac{Q^{2}}{Q-\frac{Q^{2}}{-M_{0}-\kappa ^{2}\mathcal{G}}}}.
\end{equation}%
It is solved in a closed form as%
\begin{equation}
\mathcal{G}=[\left( \kappa ^{2}+M_{0}^{2}\right) +2M_{0}Q\pm \sqrt{\mathcal{M}}]/[2\left( Q+M_{0}\right) ],
\end{equation}%
with $\mathcal{M}=Q\left( M_{0}^{2}-\kappa ^{2}\right) \left[ \left(
2Q+M_{0}\right) ^{2}-\kappa ^{2}\right] $. The impedance diverges at $\omega
=\sqrt{(\kappa -Q)/m}$. We show the impedance $Z_{00}$ for a finite free-end
chain as a function of $\omega $ in Fig.\ref{FigImpe}. It measures the
response of the plate displacement $u_{0}$ when we apply a force $f_{0}$
with frequency $\omega $.

There is a single strong impedance resonance peak in the topological phase
as shown in Fig.\ref{FigImpe}(a). It reflects the fact that the left most
plate $u_{1}$ is almost isolated in the topological phase. On the other
hand, there are many small peaks in the trivial phase as shown in Fig.\ref{FigImpe}(b). 
This is because the oscillation propagates to other plates.
The number of peaks increases as the length increases.

In passing, we provide typical values of sample parameters\cite{Mita1,Mita2}. 
They are $C=10$pF, $m=100\mu $g, $X_{\text{cap}}=100\mu $m, $\kappa =10$N$/
$m, which lead to the resonant frequency around $\omega _{0}=10$kHz. A
typical value of the damping factor is $\gamma =10^{-5}$kg$/$s. As far
as we are aware, there is no literature on the topological MEMS. 
It will be possible to construct such a MEMS with current technology\cite{MitaP}.

In this work we have revealed a novel topological dynamics of MEMS. We have
argued that it is experimentally observable in actual samples with
randomness and damping. Our results will open an attractive field of
topological MEMS.

The author is very much grateful to Y. Mita, M. Sato and N. Nagaosa for
helpful discussions on the subject. This work is supported by the
Grants-in-Aid for Scientific Research from MEXT KAKENHI (Grants No.
JP17K05490 and No. JP18H03676). This work is also supported by CREST, JST
(JPMJCR16F1 and JPMJCR20T2).

\end{document}